\documentclass{ifacconf}

\usepackage{graphicx}      % include this line if your document contains figures
\usepackage{natbib}        % required for bibliography
\usepackage{amsmath}
\usepackage{balance}

\begin{document}
\begin{frontmatter}

\title{Extended Time Varying Multi-Cluster Fluctuating Two-Ray Fading Model for Maritime Environment} 

\author[First]{Antoine Thibault Vi{\'e}} 
\author[First]{Roberto Galeazzi}
\author[First]{Dimitrios Papagergiou} 

\address[First]{Electrical Engineering Department,Technical University of Denmark, Kgs. Lyngby, Danemark, (e-mail: author@dtu.dk)}

\begin{abstract}                % Abstract of not more than 250 words.
The recent advancements in autonomous and remote operation of maritime vessels necessitates the development of robust and reliable communication systems to support high-bandwidth applications such as real-time monitoring, navigation, and control. Existing communication channel models, including Rayleigh and Rician fading, are inadequate to accurately describe the dynamic and complex nature of maritime communication, particularly for high-speed vessels in coastal environments. This paper proposes an extension to the Multi-Cluster Fluctuating Two-Ray Fading (MFTR) model that also accounts for key phenomena such as large-scale fading, time-varying parameters and Doppler shifts. The extended MFTR model integrates Stochastic Differential Equations (SDEs) to capture the time-varying characteristics of the channel, such as phase shifts and delays, while considering physical factors like delay-induced power loss and path loss. The accuracy of the proposed model is assessed in simulation.
% The proposed model provides a more comprehensive and accurate representation of the maritime communication channel, which can enhance the performance and reliability of communication systems for autonomous vessels, particularly in challenging maritime environments.
\end{abstract}


\end{frontmatter}
%===============================================================================

\section{Introduction}

The maritime industry has experienced significant growth in recent years, driven by increased global trade. The growing need for efficient and sustainable transportation solutions has heightened the interest in autonomous and remotely-piloted vessels, which promise to revolutionize maritime operations by improving safety and resilience. This shift towards high levels of automation necessitates robust and reliable communication systems capable of supporting high-bandwidth applications, such as real-time monitoring, navigation, and control of autonomous ships. Consequently, there is an increasing demand for higher data rates and more efficient communication technologies to ensure seamless connectivity in maritime environments.

For near-coast maritime operations, 4G and 5G technologies are emerging as the backbone of communication systems, offering the necessary high-speed data transfer capabilities. However, classical channel models, such as Rayleigh and Rician fading models \citep{tse2005fundamentals}, which have been widely used in wireless communication systems, are often insufficient to accurately model the dynamic and complex nature of maritime communication \citep{wangWirelessChannelModels2018}, particularly in the context of high-speed vessel operation and harsh propagation environments.

The advancements in the use of 4G and 5G wireless technologies require more sophisticated models to address the unique challenges posed by maritime environments \citep{wangWirelessChannelModels2018}, \citep{romero-jerezFluctuatingTwoRayFading2017}, \citep{hemadehMillimeterWaveCommunicationsPhysical2018}, \citep{marins2021fading}, \citep{eggers2019wireless}. In recent years, several new models have emerged that offer more accurate and flexible representations of the channel. Notably, the $\kappa-\mu$ shadowed fading distribution \citep{moreno2016kappa}, the extended Round Earth Loss (REL) model \citep{ekmanFadingDistributionModel2024}, and the Fluctuating Two-Ray (FTR) fading model \citep{romero-jerezFluctuatingTwoRayFading2016}, have all been proposed as more accurate alternatives. Furthermore, the Multi-Cluster Fluctuating Two-Ray Fading (MFTR) model \citep{sanchezMultiClusterFluctuatingTwoRay2024} has been introduced, reconciling the two dominant fading approaches, $\kappa-\mu$ shadowed fading and the FTR model, providing a simplified yet accurate representation of the channel.

Despite the advancements offered by these models, they fail to capture several important physical aspects of the communication channel. Specifically, they do not account for the large-scale fading characteristics, the time-varying nature of the channel, the inherent loss of power due to delays, Doppler shifts derived from Jakes’ model, or the impact of distance on macro parameters. These limitations reduce their accuracy in real-world scenarios, where such factors play a critical role in system performance \citep{wangWirelessChannelModels2018}.

Addressing the shortcomings of existing models will facilitate a more comprehensive and accurate representation of the maritime communication channel, enhancing the reliability of communication systems for autonomous and remote piloted vessels in near-coast environments. To this end, the paper proposes an extension to the MFTR model that includes the aforementioned physical aspects. Specifically, this work contributes by
\begin{itemize}
    \item Incorporating baseband modeling and an explicit representation of Doppler shifts.
    \item Applying Stochastic Differential Equations (SDEs) for handling time-varying parameters.
    \item Modeling of path loss effects arising from propagation delays.
\end{itemize}

The remainder of the paper is organized as follows. Sections \ref{sec:SSLargeScaleFading} and \ref{sec:SmallScaleFading} provide an overview of the large-scale and small-scale fading phenomena, respectively, prevalent in maritime communication dynamics, as well as existing modelling frameworks for them. Section \ref{sec:ModleExtenSions} details the model extensions proposed in this work. Section \ref{sec:results} demonstrates the model efficacy via simulations. Finally, Section \ref{sec:conclusions} provides concluding remarks.

\section{Near Sea-Surface Large-Scale Fading} \label{sec:SSLargeScaleFading}

Large-scale fading refers to the gradual variation of received signal power over distance due to path loss and shadowing. As mentioned in \citep{wangWirelessChannelModels2018,habibWirelessChannelModels2019,leeSeaSurfaceMobileRadiowave}, various channel models have been investigated and applied in the context of maritime communication. \citet{leeSeaSurfaceMobileRadiowave} demonstrated that certain models provide an accurate representation of real-world conditions. In our study, to derive a physically meaningful model, empirical models have been excluded, and only analytical models have been considered. Consequently, the simplest path loss model, known as Free Space Loss (FSL), is given by:
\begin{equation}
L_{FSL}(t) = \left(\frac{\lambda}{4 \pi d(t)}\right)^2 \, ,
\end{equation}
where $\lambda$ represents the signal wavelength, and $d$ denotes the transmitter-receiver distance. This model assumes an unobstructed line-of-sight (LOS) propagation environment, as well as the absence of reflections. However, in a maritime setting, multipath propagation effects must be taken into account, for example, due to signal reflections off the sea surface and nearby vessels, which can lead to time-dispersive and frequency-selective fading. This phenomenon is particularly critical for ship-to-shore and inter-ship communication links \citep{cokerMaritimeChannelModeling2013}.

As described in \citep{wangWirelessChannelModels2018}, the two-ray model provides an accurate characterization within the defined range, under the assumption of an ideal reflection coefficient $\Gamma = -1$ and a distance range of $d \in \{d_0, d_{break}\}$, where $d_{break}$ is given by:
\begin{equation}
d_{break} = \frac{4 h_t h_r}{\lambda} \, .
\end{equation}
The parameters $h_t$ and $h_r$ denote the heights of the transmitter and receiver, respectively. The corresponding path loss model is expressed as:
\begin{equation}
L_{2-ray}(t) = 4 L_{FSL}(t)\sin\left( \frac{2 \pi}{\lambda} \frac{h_t h_r}{d(t)} \right)^2 \, .
\end{equation}

For distances exceeding the breakpoint $d_{break}$, the ducting effect must be considered \citep{leeSeaSurfaceMobileRadiowave}. Therefore, when $d \in \{d_{break},d_{LOS}\}$, where $d_{LOS}$ is given by:
\begin{equation}
d_{LOS} = \sqrt{h_t^2 + 2 h_t R} + \sqrt{h_r^2 + 2h_rR}
\end{equation}
with $R$ representing the Earth's curvature, the three-ray model is defined as:
\begin{equation}
L_{3-ray} =  4L_{FSL}(t) (1+b(t))^2 \, .
\end{equation}
The parameter $b$ is given by:
\begin{equation}
b(t) = 2 \sin  \left( \frac{2 \pi h_t h_r}{\lambda d(t)}\right) \sin \left( \frac{2 \pi (h_e - h_r)(h_e - h_t)}{\lambda d(t)}\right)
\end{equation}
with $h_e$ representing the effective duct height. Consequently, the overall large-scale fading behavior can be summarized as follows:
\begin{equation}
L(t)  = \begin{cases}
L_{FSL}(t) & \text{if } d(t) < d_0, \\
L_{2-ray}(t) & \text{if } d(t) \in \{d_0,d_{break}\}, \\
L_{3-ray}(t) & \text{if } d(t) \in \{d_{break}, d_{LOS}\}
\end{cases}
\end{equation}

\section{Small-Scale Fading} \label{sec:SmallScaleFading}
Small-scale fading arises due to rapid channel variations caused by multipath propagation and Doppler effects. In maritime environments, where reflections from the sea surface and vessel motion introduce time-selective and frequency-selective fading, accurate modeling is crucial for reliable communication.  
\subsection{Channel State Preliminaries}
As explained in \citep{tse2005fundamentals} the baseband multipath channel is modeled as a linear Finite Inpulse Response (FIR) and is expressed as:
\begin{equation}
    h_b(\tau,t) = \sum_i a_i^b(t) \delta(\tau-\tau_i(t))
\end{equation}
where $a_i^b(t)$ and $\delta(\tau-\tau_i(t))$ denote amplitude of a path in the baseband and the Dirac delta function, respectively. Since the Z-transform of a delayed Dirac function is known, the channel can be written in the Z-domain as:
\begin{equation}
    H(z;t) = \sum_i a_i^b(t) z^{-\tau_i(t) f_s}
    \label{eq:channel_z_transform}
\end{equation}
where $f_s$ denotes the sampling frequency. This system is a LTV system. However, as defined in \citep{tse2005fundamentals}, the channel coherence time can be determined, representing the duration between two instances where the channel can be approximated as a linear time-invariant (LTI) system. The coherence time is given by:
\begin{equation}
    T_c = \frac{1}{4 D_s}
\end{equation}
with $D_s$ being the Doppler spread of the channel. 

\subsection{The Multi-Cluster Fluctuating Two-Ray Fading Model}\citet{romero-jerezFluctuatingTwoRayFading2017} introduced the Fluctuating Two-Ray Fading Model (FTR), and \citet{wangWirelessChannelModels2018} pointed out that it is an effective model when the distance remains below the break point. \citet{sanchezMultiClusterFluctuatingTwoRay2024} proposed an extension, which resulted in the Multi-Cluster Fluctuating Two-Ray Fading Model (MFTR), which can be applied beyond the break point by adjusting parameters when the LOS components are weaker. 

The MFTR model is expressed as:
\begin{equation}
    W = \lvert \sqrt{\zeta} \left( V_1 e^{j \phi_1} + V_2 e^{j \phi_2} \right) + Z_1 \rvert^2 + \sum_i^\mu \lvert \sqrt{\zeta} U_i e^{j \varphi_i} + Z_i \rvert^2
    \label{eq:MFTR}
\end{equation}
where $\mu$ is the number of multipath components, $U_i$ denotes the specular component of the $i^{th}$ cluster, $\phi_i,\varphi_i \sim \mathcal{U}(0,2\pi)$ are uniformly distributed random variables, $\zeta$ is a unit-mean Gamma distribution with probability density function (PDF):
\begin{equation}
    f_{\zeta}(x) = \frac{m^mx^{m-1}}{\Gamma(m)}e^{-mx} \, .
\end{equation}
Here $m$ denotes the shadowing severity index of the specular components. The LOS components $V_1,V_2$ are defined as:
\begin{equation}
    \begin{split}
        V_1 &= \sqrt{\sigma^2\mu K ( 1 + \sqrt{1-\Delta^2})} \\
        V_2 &= \sqrt{\sigma^2\mu K ( 1 - \sqrt{1-\Delta^2})}
    \end{split}
\end{equation}
where $K$ represents the ratio of the average power of the specular components to the power of the remaining scattered components, and is given by:
\begin{equation}
    K = \frac{V_1^2 + V_2^2 + \sum_i^\mu U_i^2}{2 \sigma^2 \mu}
\end{equation}

The parameter $\Delta \in [0,1]$ quantifies the similarity between the average received powers of the dominant components in cluster 1 and the power allocation to the specular components of the remaining clusters:
\begin{equation}
    \Delta = \frac{2 V_1 V_2}{V_1^2 + V_2^2 + \sum_i^\mu U_i^2}
\end{equation}

The variables $Z_i$ in~\eqref{eq:MFTR} follow a Rayleigh distribution with the parameter $\sigma$ defined as:
\begin{equation}
    \sigma = \sqrt{\frac{1}{\sqrt{2\mu(1+K)}}} \, .
\end{equation}

Hence, based on \eqref{eq:MFTR} and using \eqref{eq:channel_z_transform}, and introducing delays the channel can be expressed as: 
\begin{align}
    H_{MFTR}(z;t) &= \bigg(\sqrt{\zeta}V_1e^{j \phi_1}  + \sqrt{\zeta}V_2e^{j \phi_2} + Z_1 \bigg)z^{-\tau_1 f_s} \nonumber \\
    &\phantom{=}\, + \sum_i^\mu \bigg( \sqrt{\zeta}U_ie^{j \varphi_i} + Z_i \bigg)z^{-\tau_i f_s}
\end{align}
with $ \tau_i \sim \mathcal{HN}(0,\sigma_\tau)$ , where $\mathcal{HN}$ denotes the half-normal distribution..

\section{Proposed Extension of the Multi-Cluster Fluctuating Two-Ray Fading Model} \label{sec:ModleExtenSions}
%While the MFTR fading model effectively captures clustered multipath and fluctuating line-of-sight conditions, it falls short in maritime applications due to its limited ability to account for time-varying sea surface reflections, dynamic vessel-induced shadowing, and long-delay multipath components with associated power loss, phenomena that are characteristic of over-water propagation.
%To overcome the above listed shortcomings of the MFTR we suggest to extend this model by incorporating baseband modeling, Doppler shifts and Stochastic Differential Equations (SDEs) for time-varying parameters. 
%The baseband model provides a tractable framework for simulating and analyzing the complex envelope. Doppler shifts account for relative motion between the transmitter, receiver, and scatterers. Stochastic differential equations (SDEs) introduce realistic temporal dynamics beyond static or simplified assumptions. Finally, incorporating associated power enables tracking of delay-dependent attenuation in scattered components, capturing the physical power loss over extended multipath paths.
While the MFTR fading model captures clustered multipath and fluctuating line-of-sight conditions, it remains insufficient for maritime scenarios, where time-varying sea surface reflections, vessel-induced shadowing, and long-delay components with associated power loss are prominent. To address these limitations, we propose an extended model that incorporates baseband representation, Doppler effects, and stochastic differential equations (SDEs) for time-varying parameters. The baseband formulation offers a tractable framework for complex envelope analysis; Doppler shifts capture relative motion dynamics; and SDEs introduce realistic temporal evolution. Additionally, modeling associated power allows delay-dependent attenuation to be tracked, reflecting physical path loss over extended multipath trajectories.

\subsection{Baseband \& Doppler Shift effects}
The previous model did not incorporate the carrier frequency of the signal or the effects of Doppler shifts. These factors can be integrated into the model, leading to the following updated formulation: 
\begin{equation}
\begin{split}
    H(z;t) =& \bigg( \sqrt{\zeta}V_1e^{j(\phi_1 - \omega_c \tau_1 + \omega_{d_1}t)}  \\ &+ \sqrt{\zeta}V_2e^{j (\phi_2 - \omega_c \tau_1 + \omega_{d_2}t)} + Z_1 \bigg)z^{-\tau_1 f_s} \\ &+ \sum_i^\mu  \bigg( \sqrt{\zeta}U_ie^{j (\varphi_i -\omega_c \tau_i + \omega_{d_i}t)} + Z_i \bigg) z^{-\tau_i f_s}
\end{split}
\end{equation}
where
\begin{equation}
    Z_i = \sigma \sqrt{X_i^2 + Y_i^2} e^{-j (\varphi_{z_i} - \omega_c \tau_{i} + \omega_{d_i}t)}, \text{ } i = \{1, \dots, \mu\}
\end{equation}
and $\omega_c = 2\pi f_c, \omega_{d_i} = 2\pi f_{d_i}$, with $f_c$ the carrier frequency and $f_{d_i}$ the Doppler frequency of the $i^{th}$ cluster.

\subsection{Time dependency of channel parameters}
The parameters in wireless communication channels are inherently time-varying \citep{fengStochasticDifferentialEquation2007,olamaStochasticDifferentialEquations2009,mossbergStochasticDifferentialEquation2009}. Consequently, Stochastic Differential Equations (SDEs) are employed to characterize their temporal variations and to accurately model their statistical distributions.

Following the formulation in \citep{mossbergStochasticDifferentialEquation2009} the phase increment for the $i^{th}$ cluster is defined as:
\begin{equation}
    d \phi_i(t) = \sqrt{C_i^{\phi}} dW_i^{\phi}(t)
\end{equation}
where $C_i^{\phi}$ denotes the variance of the process and $dW_i^{\phi}(t)$ is the increment of the Wiener process $W_i^\phi(t)$. The Wiener processes of the phase increments are assumed to be independent.

The time dependency of the delays is modeled as:
\begin{equation}
    d \tau_i(t) = -\tau_i(t) dt + \sqrt{2C_i^\tau} dW_i^{\tau}(t) \text{ s.t } \tau_i(t) \geq 0
\end{equation}
with $C_i^\tau$ the variance and $dW_i^{\tau}(t)$ the increment of the Wiener process $W_i^{\tau}(t)$. The same independence assumption as for the phase processes is applied to the Wiener processes associated with the delays.

The Rayleigh-distributed amplitude is defined as: 
\begin{equation}
    \alpha_i(t) = \sigma \sqrt{X_i(t)^2+Y_i(t)^2}
\end{equation}
where $X_i(t), Y_i(t)$ are designed to approximate a normal distribution. Their increment are given by:
\begin{equation}
    \begin{split}
        dX_i(t) &= -X_i(t) dt + \sqrt{2}dW_i^X (t)\\
        dY_i(t) &= -Y_i(t) dt + \sqrt{2}dW_i^Y(t)
    \end{split}
\end{equation}
where $dW_i^X (t)$ and $dW_i^Y(t)$ are the increments of the corresponding Wiener processes, which are also assumed to be independent. Consequently, the multipath components are expressed as:
\begin{equation}
    Z_i(t) = \alpha_i(t) e^{j(\phi_i(t) -\omega_c \tau_i(t) + \omega_d t)} \, .
\end{equation}

Regarding the shadowing component, the unit-mean $\Gamma$ distribution is approximated using the Cox-Ingersoll-Ross model in order to introduce realistic temporal dynamics into the shadowing process: 
\begin{equation}
    d\zeta(t) = m(1 - \zeta(t))dt + \sqrt{2\lvert \zeta(t) \rvert} dW^\zeta(t)
\end{equation}
where $dW^\zeta(t)$ is the increment of the Wiener process $W^\zeta(t)$. Finally, the variables and their increments are represented in vector form as:

\begin{equation}
    \Lambda(t) = \begin{bmatrix}
        \zeta(t) \\ \phi_{1}(t) \\ \phi_{2}(t) \\ \tau_{1}(t) \\ \begin{bmatrix}
            X_{i}(t) \\ Y_{i}(t) \\ \varphi_{i}(t) \\ \tau_{i}(t)
        \end{bmatrix}_\mu
    \end{bmatrix}, \quad d\Lambda(t) = \begin{bmatrix}
        d\zeta(t) \\ d\phi_{1}(t) \\ d\phi_{2}(t) \\ d\tau_{1}(t) \\ \begin{bmatrix}
           d X_{i}(t) \\ dY_{i}(t) \\ d\varphi_{i}(t) \\ d\tau_{i}(t)
        \end{bmatrix}_\mu
    \end{bmatrix}
\end{equation}

\subsection{Physics Considerations}
Several physical considerations are incorporated to enhance the accuracy of the model. Firstly, a longer delay corresponds to a greater traveled distance for the associated cluster. Consequently, a power loss is introduced as a function of the delay. The formulation of this loss is expressed as:
\begin{equation}
    L_i(t) = L_i(d(t),\tau_i(t)) = \left( \frac{d(t)}{d(t) + c\tau_i(t)}\right)^2
\end{equation}
where $c$ is the speed of light.

Secondly, given the distance, the minimum delay is expressed as $\tau_{min} = \frac{d}{c}$ and is established as a factor for the channel. Consequently, the remaining delays represent the delay spread. %Finally, the path loss as a function of distance is considered, leading to the following expression for the channel:
Finally, the Extended Time-Varying Multi-Cluster Fluctuating Two-Ray (ETVMFTR) fading channel is formulated as:
\begin{equation}
    \begin{split}
       H(z;t) =& \sqrt{L(t)L_1(t)}z^{-\left( \tau_1 + \frac{d(t)}{c}\right)f_s} \\&\bigg(\sqrt{\zeta(t)}V_1 e^{j(\phi_1(t) - \omega_c \tau_1(t) + \omega_{d_1}t)}\\
        &+\sqrt{\zeta(t)} V_2 e^{j (\phi_2(t) - \omega_c \tau_1(t) + \omega_{d_2}t)} + Z_1(t) \bigg) \\ &+ 
    \sqrt{L(t)} z^{-\frac{d(t)}{c} f_s} \\& \bigg(\sum_i^\mu \Big(\sqrt{L_i(t)\zeta(t)} U_i e^{j (\varphi_i(t) -\omega_c \tau_i(t) + \omega_{d_i}t)} \\&+ \sqrt{L_{i}(t)}{}Z_i(t) \Big)z^{-\tau_i(t) f_s} \bigg)
    \end{split}
\end{equation}
where
\begin{equation}
    Z_i(t) = \sigma \sqrt{X_i(t)^2+Y_i(t)^2}e^{-j (\varphi_i(t) - \omega_c \tau_i(t) + \omega_{d_i}t)} \, .
\end{equation}

To ensure consistency in the total received power across different channel realizations and to prevent unintended biases in system performance evaluation, power normalization is applied:
\begin{equation}
    H_{n}(z;t) = \frac{H(z;t)}{\sqrt{V_1^2 + V_2^2 + 2\mu\sigma^2}}
\end{equation}

Additionally, it is assumed that the LOS components experience a lower delay than the scattered components:
\begin{equation}
    \tau_1 \leq \{\tau\}_\mu
\end{equation}
where $\{\tau\}_\mu$ denotes the set of $\mu$ delays.

\section{Simulation results} \label{sec:results}
In this section, the performance of the proposed channel model is analyzed through numerical simulations. The statistical properties of the modeled channel, including delay spread, Doppler effects, and power variations, are examined to validate its accuracy. Further, the impact of key parameters on channel behavior is evaluated to demonstrate the model’s reliability in realistic wireless communication scenarios.

\begin{figure}[h!]
    \centering
    \includegraphics[width=6.4cm]{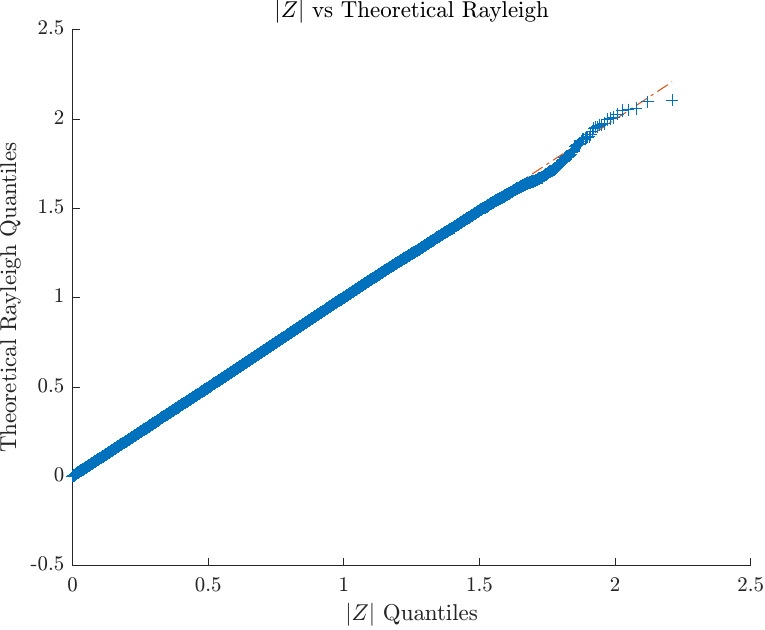}
    \caption{$\lvert Z \rvert$ quantiles comparison to Rayleigh Distribution theoretical quantiles.}
    \label{fig:z_qqplot}
\end{figure}

\begin{figure}[t]
    \centering
    \includegraphics[width=6.4cm]{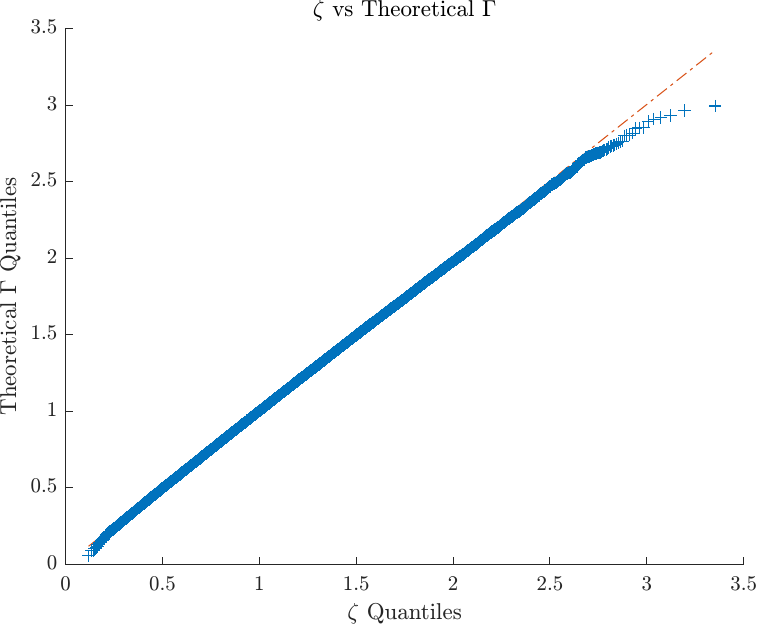}
    \caption{$\zeta$ quantiles comparison to $\Gamma$ Distribution theoretical quantiles.}
    \label{fig:zeta_qqplot}
\end{figure}

Firstly, the statistical behavior of the different parameters is analyzed using quantiles, the correlation factor ($R^2$) and the Mean-Square Error (MSE) are presented in the Table~\ref{tb:margins}:

\begin{table}[h!]
\begin{center}
\caption{Quantile-based variable correlation to predefined distribution.}\label{tb:margins}
\begin{tabular}{c|c|c}
Variable & $R^2$ & MSE  \\\hline
$\zeta$ & $0.9993$ & $9.9962e-^{5}$ \\
$X$ & $0.9993$ & $8.5040e^{-4}$  \\ 
$Y$ & $0.9992$ & $9.2193e^{-4}$  \\ 
$\tau_i$ & $0.9987$ & $4.7114e^{-16}$ \\
$\lvert Z \rvert $ & $ 0.9998$ & $1.6018e^{-5}$ \\
\hline
\end{tabular}
\end{center}
\end{table}

where $\tau_i$ now follows a Weibull distribution due to the non-negativity of the propagation delays, as the absolute value operation guarantees physically meaningful, strictly positive delay values, and $\tau_1$ does not follow any specific distribution since it is the minimum of $\mu+1$ SDE. The comparisons of the quantiles to their respective theoretical distributions for $\lvert Z \rvert$ and $\zeta$ are presented in Fig.\ref{fig:z_qqplot} and Fig.\ref{fig:zeta_qqplot}, respectively, thereby illustrating their closeness to the expected distributions.

% \begin{figure}[h!]
%     \centering
%     \includegraphics[width=8.4cm]{Results/z_histogram.pdf}
%     \caption{$\lvert Z \rvert$ distribution compared to the theoretical Rayleigh distribution}
%     \label{fig:z_histogram}
% \end{figure}

% \begin{figure}[h!]
%     \centering
%     \includegraphics[width=8.4cm]{Results/zeta_histogram.pdf}
%     \caption{$\zeta$ distribution compared to the theoretical $\Gamma$ distribution}
%     \label{fig:zeta_histogram}
% \end{figure}

The evolution of the channel over distance and time is simulated under the assumption of perfect channel knowledge and the application of a zero-forcing equalizer. A Quadrature Amplitude Modulation (QAM) 16 scheme and an Orthogonal frequency-division multiplexing (OFDM) system with 1024 subcarriers and 1024/16 guard bands are used. The simulation utilizes the parameters listed in Table~\ref{tab:antennas_ship_param} where $G_r,G_t$ are the antenna receiver and transmitter gains, $P_t$ is the transmitted power, $P_w$ is the noise power and $v$ is the speed of the ship.  To simplify the simulation, all components are assumed to share the same Doppler frequency and variance.

\begin{table}[htbp]
    \centering
    \caption{Simulation parameters.}
    \begin{tabular}{c|c}
        Variables & Value \\ \hline
         $K, \Delta, \mu, m$ & 10.788, 0.29, 40, 90.252 \\
         $C_\phi, C_\tau$ & $2 \pi, 1e^{-10}$ \\
         $f_c, f_d, f_s$ & 5 GHz, 100 Hz, 20 MHz \\
         $T_c$ & 10 ms \\
         $G_r, G_t$ & 0 dB, 0 dB \\
         $P_t, P_w$ & 40 dBm, -100 dBm \\
         $h_t, h_r, h_e$ & 8 m, 15 m, 35 m \\
         $v, d_0$ & 25 km/h, 200 m \\
         \hline
    \end{tabular}    
    \label{tab:antennas_ship_param}
\end{table}

\begin{figure}[tbp]
    \centering
    \includegraphics[width=6.4cm]{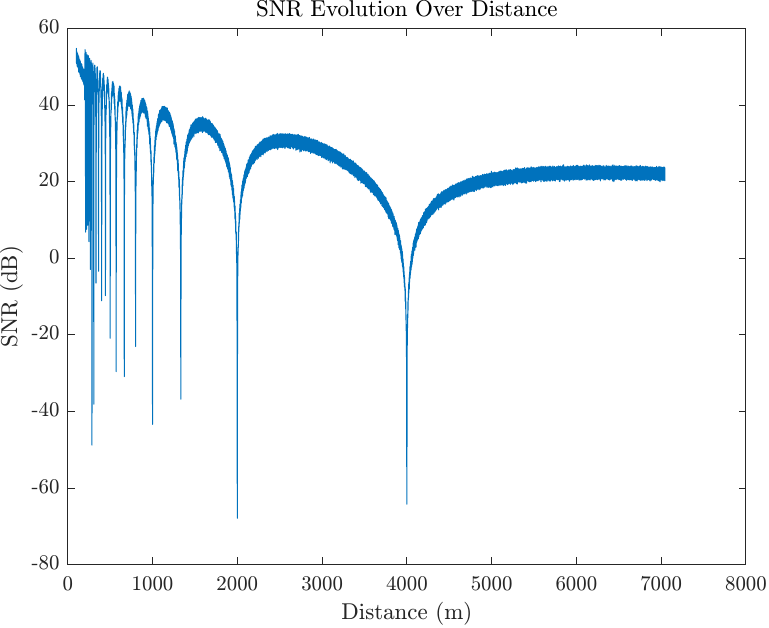}
    \caption{SNR evolution over distance for the parameters given in Table~\ref{tab:antennas_ship_param}.}
    \label{fig:SNR_Evolution}
\end{figure}

\begin{figure}[tbp]
    \centering
    \includegraphics[width=6.4cm]{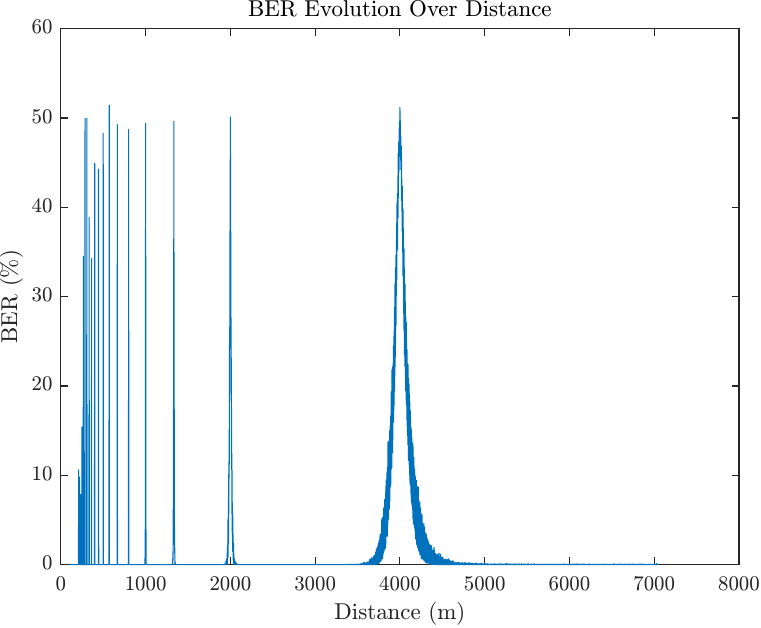}
    \caption{BER evolution over distance for the parameters given in Table~\ref{tab:antennas_ship_param}.}
    \label{fig:BER_Evolution}
\end{figure}

\begin{figure}[tbp]
    \centering
    \includegraphics[width=6.4cm]{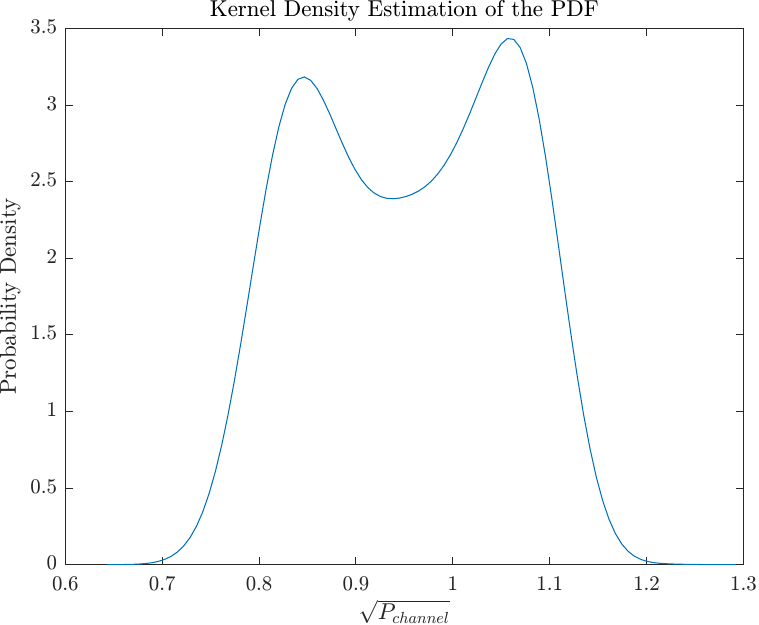}
    \caption{Empirical PDF for the parameters given in table~\ref{tab:antennas_ship_param}.}
    \label{fig:PDF}
\end{figure}

\begin{figure}[h!]
    \centering
    \includegraphics[width=6.4cm]{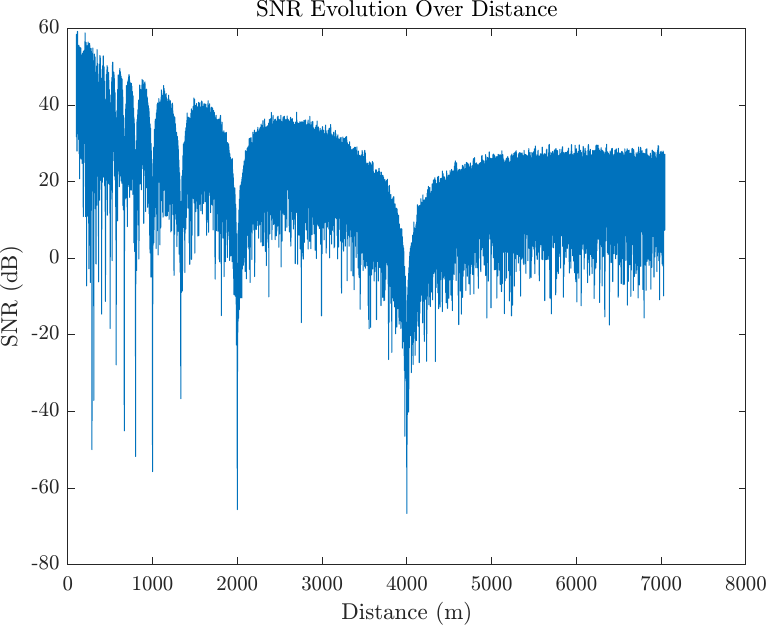}
    \caption{SNR evolution over distance for the parameters given in Table~\ref{tab:antennas_ship_param} with $K=4.225,\Delta=0.999,\mu=1,m=38.868$.}
    \label{fig:SNR2_Evolution}
\end{figure}

\begin{figure}[h!]
    \centering
    \includegraphics[width=6.4cm]{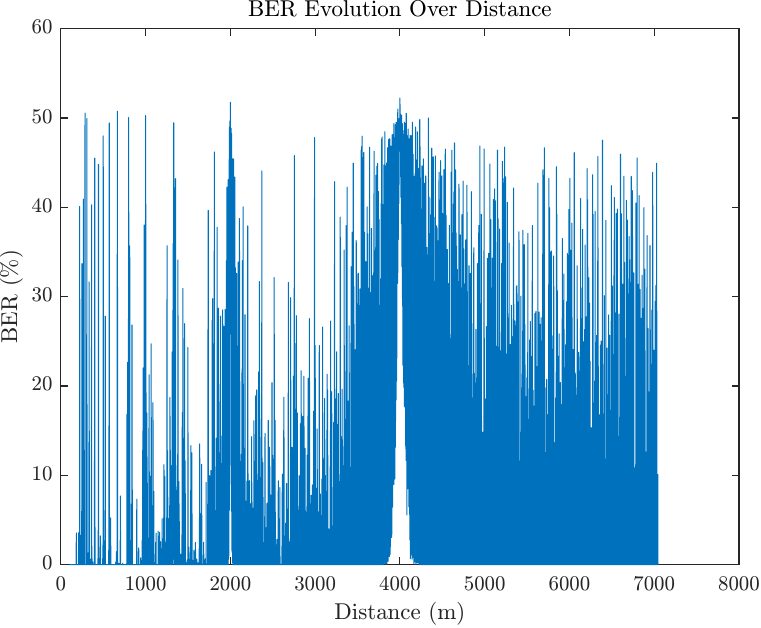}
    \caption{BER evolution over distance for the parameters given in Table~\ref{tab:antennas_ship_param} with $K=4.225,\Delta=0.999,\mu=1,m=38.868$.}
    \label{fig:BER2_Evolution}
\end{figure}

\begin{figure}[h!]
    \centering
    \includegraphics[width=6.4cm]{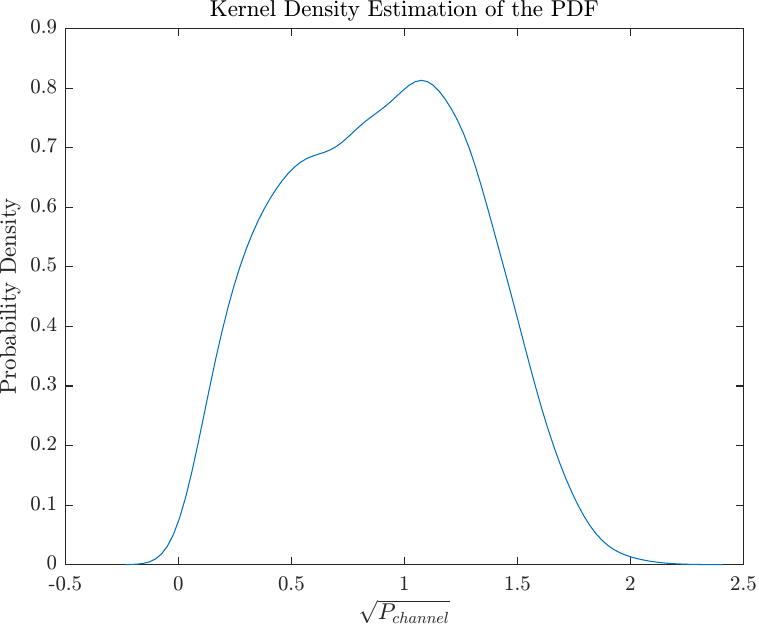}
    \caption{Empirical PDF for the parameters given in Table~\ref{tab:antennas_ship_param} with $K=4.225,\Delta=0.999,\mu=1,m=38.868$.}
    \label{fig:PDF2}
\end{figure}

The evolution of the Signal-to-Noise Ratio (SNR) across distance is illustrated in Fig.~\ref{fig:SNR_Evolution}.
It can be observed that the channel may enter a deep fading state depending on the distance, especially when two or three LOS components are in phase opposition. Additionally, small-scale fading introduces an envelope over large-scale fading variations. As a result, the Bit-Error Ratio (BER) fluctuates accordingly, as shown in Fig.~\ref{fig:BER_Evolution}.
In the case of deep fading, the channel becomes unusable due to a very low SNR, resulting in a significantly high BER. The PDF of the simulated channel is illustrated in Fig.~\ref{fig:PDF}.
The PDF aligns well with the measured LOS data presented in Fig. 5a of \citep{sanchezMultiClusterFluctuatingTwoRay2024}.
A similar analysis can be performed using different macro-parameters $K,\Delta,\mu$ and $m$. 
Figures~\ref{fig:SNR2_Evolution}, \ref{fig:BER2_Evolution}, and \ref{fig:PDF2} illustrate that degraded channel conditions are being experienced, leading to a reduction in average SNR and an increased frequency of instances where the channel becomes unusable. As previously observed, the probability density function is in close agreement with the measured LOS conditions reported in Fig.~5b of \citep{sanchezMultiClusterFluctuatingTwoRay2024}.

\section{Conclusions and future Work} \label{sec:conclusions}

This work presented an extended multi-cluster two-ray fading model that describes communication channels in maritime operation. The integration of a baseband model, a description of Doppler shifts, and the inclusion of Stochastic Differential Equations (SDEs) for handling time-varying parameters allowed for an improved model accuracy. Numerical simulations validated the theoretical findings.
Future work includes deriving distance-dependent estimation laws for $K$,$\Delta$ and $\mu$ incorporating transmitter height dynamics, and experimentally validating the proposed model.
%Future work will include the derivation of an estimation law for the parameters $K$,$\Delta$ and $\mu$ as a function of distance, the consideration of the transmitter height dynamics, and experimental verification of the modeling framework.

\section*{Acknowledgments}
This research was conducted at DTU Electro as part of the SLGreen project. Shipping Lab and SLGreen is
partly founded by Innovation Fund Denmark (IFD) under File No. 8090-00063B and by Danish Maritime Fund, Lauritzen Fonden, Orient’s Fond and DS Norden. The authors appreciate their support.

\balance
\bibliography{references}             % bib file to produce the bibliography
                                                     % with bibtex (preferred)
                                                   
%\begin{thebibliography}{xx}  % you can also add the bibliography by hand

%\bibitem[Able(1956)]{Abl:56}
%B.C. Able.
%\newblock Nucleic acid content of microscope.
%\newblock \emph{Nature}, 135:\penalty0 7--9, 1956.

%\bibitem[Able et~al.(1954)Able, Tagg, and Rush]{AbTaRu:54}
%B.C. Able, R.A. Tagg, and M.~Rush.
%\newblock Enzyme-catalyzed cellular transanimations.
%\newblock In A.F. Round, editor, \emph{Advances in Enzymology}, volume~2, pages
%  125--247. Academic Press, New York, 3rd edition, 1954.

%\bibitem[Keohane(1958)]{Keo:58}
%\newblock \emph{Power and Interdependence: World Politics in Transitions}.
%\newblock Little, Brown \& Co., Boston, 1958.

%\bibitem[Powers(1985)]{Pow:85}
%\newblock Is there a way out?
%\newblock \emph{Harpers}, pages 35--47, June 1985.

%\bibitem[Soukhanov(1992)]{Heritage:92}
%\newblock \emph{{The American Heritage. Dictionary of the American Language}}.
%\newblock Houghton Mifflin Company, 1992.

%\end{thebibliography}

\end{document}